\documentclass[prx,aps,twocolumn,superscriptaddress,longbibliography]{revtex4-2}

\usepackage[pdftex]{graphicx}
\usepackage[dvips]{epsfig}
\usepackage{float}
\usepackage{subfigure}

\usepackage{amsbsy,amssymb,amsmath,bm,mathtools}

\usepackage{xspace}
\usepackage{bm}
\usepackage{color}
\usepackage{makecell}

\usepackage{url}
\usepackage[section]{placeins}
\usepackage[colorlinks=true,linkcolor=blue,citecolor=blue]{hyperref}

\begin{document}



\title{Machine learning approach for vibronically renormalized electronic band structures}

\author{Niraj Aryal }\email{naryal@bnl.gov}
\affiliation{Condensed Matter Physics and Materials Science Division, Brookhaven National Laboratory, Upton, New York 11973, USA}

\author{Sheng Zhang}
\thanks{The first two authors contributed equally to this work.}
\affiliation{Department of Physics, University of Virginia, Charlottesville, VA 22904, USA}

\author{Weiguo Yin}
\affiliation{Condensed Matter Physics and Materials Science Division, Brookhaven National Laboratory, Upton, New York 11973, USA}

\author{Gia-Wei Chern}\email{gchern@virginia.edu}
\affiliation{Department of Physics, University of Virginia, Charlottesville, VA 22904, USA}

\date{\today}

\begin{abstract}
We present a machine learning (ML) method for efficient computation of vibrational thermal expectation values of physical properties from first principles. Our approach is based on the non-perturbative frozen phonon formulation in which stochastic Monte Carlo algorithm is employed to sample configurations of nuclei in a supercell at finite temperatures based on a first-principles phonon model. A deep-learning neural network is trained to accurately predict physical properties associated with sampled phonon configurations, thus bypassing the time-consuming {\em ab initio} calculations. To incorporate the point-group symmetry of the electronic system into the ML model, group-theoretical methods are used to develop a symmetry-invariant descriptor for phonon configurations in the supercell.  We apply our ML approach to compute the temperature dependent electronic energy gap of silicon based on density functional theory (DFT). We show that, with less than a hundred DFT calculations for training the neural network model, an order of magnitude larger number of sampling can be achieved for the computation of the vibrational thermal expectation values. Our work highlights the promising potential of ML techniques for finite temperature first-principles electronic structure methods. 
\end{abstract}

\maketitle

\section{Introduction}

\label{sec:intro}

Recent advances in electronic structure methods and rapid progress in computational capabilities and artificial intelligence techniques have enabled accurate and fast computation of materials’ properties. This new paradigm of materials research is further assisted by the creation of large freely available databases containing many years worth of human knowledge~\cite{jain_apl2013,scaling_deeplearning_Merchant_Nature2023}. 
For example, several machine learning (ML) models have been developed for accurate and efficient structure-property mapping from large databases of Kohn-Sham density functional theory (DFT) calculations~\cite{Hansen13,Schutt14,Hansen15,Schutt17}. Another prominent application is the ML based force-field or interatomic potential models trained by dataset from DFT calculations~\cite{behler07,bartok10,li15,behler16,shapeev16,botu17,smith17,zhang18,deringer19,chmiela17,chmiela18}. Such ML models, which are essentially classical force-field models, yet with a desired quantum accuracy, allows for larger scale and longer time {\em ab initio} molecular dynamics simulations. More fundamentally, ML models are also shown to provide accurate approximations of the density functionals or the Hohenberg-Kohn mapping from external potential to electron density functionals~\cite{Snyder12,Brockherde17,Nagai20,Li21,James21,Pederson22}.

The integration of ML and data science techniques with {\em ab initio} electronic structure methods have provided a tantalizing prospect of inverse materials design where a novel material of a given functionality can be predicted from available experimental measurements and theoretical calculations~\cite{sekoPRL15,FaberPRL16,xue_natcom16,curtarolo_natmat2013}. Yet, despite tremendous progress, efficient calculation of materials' properties beyond the idealized zero temperature  remains a challenge for a successful data-driven design and discovery pipeline. In particular, one important thermal effect is the phonon-induced renormalization of electronic structures~\cite{Giustino17,Shang23}. This renormalization is the main mechanism for the temperature dependence of band gap energy.

A well-developed first-principles approach to incorporate electron-phonon coupling is based on the density functional perturbation theory (DFPT)~\cite{Baroni01,Giustino07,Bernardi16}. For example, both phonon dispersion relations and electron-phonon matrix elements can be obtained from DFPT. These calculations can then be combined with the perturbative Feynman diagram methods to compute, e.g. the phonon-induced electron self-energies, including both the Fan and the Debye-Waller terms~\cite{Fan51,allen76,allen81,Cardona05}. A consistent theory of temperature-dependent band structures up to second-order in electron-phonon coupling was developed by Allen, Heine and Cardona (AHC)~\cite{allen76,allen81}. Various phonon-induced thermal effects, such as band gap renormalization and broadening of absorption edge, can be included within this framework~\cite{Giustino10,Kawai14,Ponce14}. The perturbative calculation of the electron-phonon interaction can also be elegantly integrated with the $GW$ approximation~\cite{Hedin65,Aryasetiawan98,Onida02} to partially account for electron-electron interactions. However, as this first-principles approach is built on the Kohn-Sham DFT, it is also restricted to this particular electronic structure method. 

The frozen phonon method~\cite{Yin80,Lam82,Dacorogna85,Fleszar85,Lam86,Capaz05,Monserrat18} offers an alternative framework for first-principles electron-phonon calculations. This approach is also more general in the sense that it can be used with any electronic structure solver, which implies that systems with weak and strong electronic correlations can be treated on the same footing. In fact, one of the earliest {\em ab initio} electron-phonon calculations was based on the frozen phonon method~\cite{Dacorogna85}. The central idea of this approach is rather straightforward: by comparing electronic structure solutions without and with the atomic displacements of a certain normal mode of the lattice, one can numerically compute the shift in electron eigen-energies or other physical observables induced by this particular phonon modes. This numerical calculation, often based on the finite difference method up to quadratic order, is then repeated for all phonon modes allowed in a supercell~\cite{Williams15,Monserrat18}.

Alternatively, thermal effects of lattice fluctuations can be directly computed by Monte Carlo sampling of thermal phonon configurations~\cite{patrick13,zacharias15,monserrat14,Monserrat15}. Each sampled configuration of the frozen nuclei defines an electronic structure problem with frozen atomic displacements. Expectation values of physical quantities, such as band gap, at finite temperatures are obtained by averaging over solutions of the sampled configurations.  Instead of Monte Carlo sampling, dynamical methods such as adiabatic {\em ab initio} molecular dynamics (MD)~\cite{Alder59,Car85,Pan14} and path-integral MD~\cite{Cao94a,Cao94b,Ramirez06,Morales13} have also been used to generate the phonon configurations. The vibrational expectation value of physical quantities here is obtained by averaging along dynamical paths.

Compared with the above finite-difference frozen phonon method, the approach of sampling phonon configurations, either stochastically or dynamically, has the advantage that higher-order electron-phonon couplings can be straightforwardly included~\cite{monserrat14,Monserrat15}. However, the sampling approach is computationally very demanding as an electronic structure problem has to be solved for each sampled frozen phonon configuration in a supercell, and accurate evaluation of physical quantities requires a large number of configurations. To circumvent this computational difficulty, it was shown that the configurational averaging can be approximated by averaging over a set of special configurations called thermal lines~\cite{monserrat16,monserrat16b} or even by fully deterministic supercell calculations based on a single optimal configuration of the atomic positions~\cite{zacharias16}. While these approximations significantly reduce the cost of repeated electronic structure calculations, they are exact only in the thermodynamic limit of large supercells. In practical implementations of finite supercells, their accuracy has to be carefully benchmarked.

In this work, we propose a machine learning (ML) approach to solve the efficiency issue of configurational averaging method.  The central idea is to build a ML model that accurately predicts the physical quantities of interest corresponding to a configuration of nuclei in a supercell. Monte Carlo (MC) algorithm based on an {\em ab initio} phonon model is used to sample atomic configurations, and the trained ML model is employed to efficiently predict the corresponding physical property. As a proof of principle, we apply this MC-ML approach to compute the temperature dependent electronic band gap of silicon crystal. We use silicon as an example as it has been extensively studied by using various methods and thus provides a testbed for validating our methods. Our approach not only produces accurate temperature dependence of the gap energy, but also significantly reduces the number of DFT calculations in the process. 

The ML model in our proposed framework can be viewed as a special case of the general ML structure-property mapping models~\cite{Jung19,Fan20,Hundi19,Zhang20,Rasul21} which have played a central role in the high-throughput materials design strategy. A proper descriptor of the input structure is an important ingredient of such ML models. In our case, the descriptor is particularly important for preserving the symmetry properties of the original electronic model in the mapping from phonons to properties. To this end, we employ group-theoretical methods to obtain generalized coordinates of atomic configurations in a supercell which are invariant under the point group symmetry of the crystal. The use of the phonon descriptor enhances both the training efficiency and the accuracy of the ML model. 

The rest of the paper is organized as follows. The stochastic approach to compute vibrational averages and the ML model that maps the phonon configuration to physical properties are discussed in Sec.~\ref{sec:framework}. Details of the group-theoretical methods for the phonon descriptor is presented in Sec.~\ref{sec:descriptor}. The approach is then applied to compute the temperature dependent energy gap of silicon crystal in Sec.~\ref{sec:results}. Finally, we conclude the paper and present an outlook in Sec.~\ref{sec:conclusion}.

\section{Stochastic Formalism and machine learning models}
\label{sec:framework} 

Our framework of vibronically renormalized electronic band structures is based on the stochastic frozen-phonon method. There are two major components in our framework: (i)~the {\em ab initio} lattice dynamics model for sampling the phonon configurations in a supercell, and (ii)~the ML model that maps the phonon configuration to the physical properties of interest, e.g. the electronic band gap energy. 


\subsection{{\em ab initio} phonon models}

\label{sec:ab-initio-phonon}

The fundamental assumption of most {\em ab initio} electron-phonon calculations, especially the frozen-phonon methods, is the Born-Oppenheimer (BO) or adiabatic approximation. The fact that masses of nuclei are much larger than that of electron indicates that the electron velocities are much larger than the nuclear ones, suggesting that electrons can follow the motion of the slow nuclei almost instantaneously. The well-separated time scales of the two sets of degrees of freedom allows one to write the full nuclei-electron wave function as a product form $|\Psi \rangle = | \chi  \rangle \otimes |\psi \rangle$,  where $|\chi  \rangle$ and $|\psi  \rangle$ are the phonon and electron wave functions, respectively. Under the adiabatic approximation, one can integrate out the electron degrees of freedom to obtain a potential energy surface $\mathcal{E}(\mathbf R_1, \mathbf R_2, \cdots, \mathbf R_N)$ as a function of the coordinates $\mathbf R_i$ of the nuclei. Practically, this involves solving an electronic structure problem with each atomic configuration frozen in $\mathbf R_i$ based on DFT calculations, or other electronic structure methods.

The Hamiltonian of nuclei in BO approximation is given by $\mathcal{H}_{\rm nucl} = \sum_i \mathbf P_i^2 / 2 M_i + \mathcal{E}(\{\mathbf R_i\})$, where $\mathbf P_i$ is the momentum operator of the $i$-th nuclei and $M_i$ is the corresponding mass. However, since the computation of the potential energy for each atomic configuration $\{\mathbf R_i\}$ requires solving an electronic structure problem, direct {\em ab initio} determination of the phonon modes  is computationally intractable, if not impossible. As discussed in Sec.~\ref{sec:intro}, ML force-field approaches offer an efficient and accurate approximation to the potential energy surface.  However, due to the lack of explicit analytical expressions for $\mathcal{E}(\{\mathbf R_i\})$, the phonon modes cannot be directly derived from the ML model. Vibrational thermal average has to be done in conjunction with dynamical simulations based on the ML force-field models. 

Direct phonon or lattice models can be obtained based on the harmonic  approximation to the nuclear Hamiltonian $\mathcal{H}_{\rm nucl}$. For crystalline systems, nuclei only undergo small amplitude oscillations about their equilibrium positions. We can then write the nuclear positions as $\mathbf R_i = \mathbf R^{(0)}_i + \mathbf u_i$, where $\mathbf R^{(0)}_i$ is the equilibrium position and $\mathbf u_i$ is the displacement vector. Assuming small displacements, one can then express the potential energy $\mathcal{E}(\{\mathbf R_i\})$ in a power series of the displacement vectors. Since the equilibrium positions minimize the potential energy, the linear term in the expansion vanishes. To the leading second-order, the potential energy is $\mathcal{E} = \mathcal{E}_0 + \sum_{ij} \sum_{\alpha, \beta} D_{i\alpha, j\beta} u_i^\alpha u_j^\beta$, where $u_i^\alpha$ denotes the $\alpha$-component of the displacement vector of the $i$-th nucleus ($\alpha = x, y, $ and $z$), and $D_{i\alpha, j\beta} = \partial^2 \mathcal{E} / \partial u_{i, \alpha} \partial u_{j, \beta}$, evaluated at the equilibrium positions $\{\mathbf R^{(0)}\}$, is the dynamical matrix which encapsulates the normal modes of the lattice. 

The normal modes $\mathbf V^{(\nu)}_{s, \mathbf k}$ of the crystalline systems are obtained by diagonalizing the harmonic Hamiltonian, here $\nu$ is the branch-index, $s$ is the sublattice index, and $\mathbf k$ is the reciprocal lattice wave vector. The displacement vectors can then be expressed as
\begin{eqnarray}
	\label{eq:normal_modes}
	\mathbf u_i = \sum_{\nu, \mathbf k }  \xi_{\nu, \mathbf k} \, \mathbf V^{(\nu)}_{s_i, \mathbf k} e^{i \mathbf k \cdot \mathbf R^{(0)}_i},
\end{eqnarray}
where $\xi_{\nu, \mathbf k}$ are the effective or normal-mode coordinates. In terms of the effective coordinates, the vibrational dynamics of the lattice is described by the following phonon Hamiltonian
\begin{eqnarray}
	\label{eq:H_phonon}
	\mathcal{H}_{\rm phonon} =  \sum_{\nu}\sum_{\mathbf k} \left( \frac{-\hbar^2}{2} \frac{\partial^2}{\partial \xi_{\nu, \mathbf k}^2}
	+ \frac{1}{2} \Omega_{\nu, \mathbf k}^2 \xi_{\nu, \mathbf k}^2 \right).
\end{eqnarray}
Here $\Omega_{\nu {\bf k}}$ is the eigen-frequency of the normal modes. 
{\em Ab initio} calculation of the phonon modes can now be routinely computed using DFT methods, both based on the DFPT or frozen-phonon formalisms. Computation of the dynamical matrices using methods beyond DFT have also been demonstrated~\cite{GWPT_Louie2019,DMFT_phonons_Bartomeu2020}.


\subsection{Stochastic vibrational average}

Next we consider the computation of observables such as  in the frozen-phonon framework. Quantum mechanically, an observable corresponds to a Hermitian operator $\mathcal{O}(\{\mathbf r_l\}, \{\mathbf R_i\})$ which depends on both electron and nuclei coordinates. By integrating out the electronic degrees of freedom within the Born-Oppenheimer approximation, the observable 
\[
	\mathcal{O}(\{\mathbf{R}_i\}) = \langle \psi(\{\mathbf{r}_l \})| \mathcal{O}(\{\mathbf{r}_l\}, \{\mathbf{R}_i \}) |\psi( \{\mathbf r_l\} ) \rangle
\]
becomes a function of nuclear coordinates only. The electron quantum state $|\psi( \{\mathbf r_l\} ) \rangle$ in general also implicitly depends on the nuclei configuration $\{\mathbf R_i \}$. The vibrational expectation value of the observable quantity at finite temperature $T$ is then given by~\cite{Monserrat18}
\begin{eqnarray}
   \langle \mathcal{O(T)}\rangle = \frac{1}{\mathcal{Z}} \sum_{M}  \langle \chi_M(\{\mathbf{R}_i\})| \mathcal{O}(\{\mathbf{R}_i\})|\chi_M(\{\mathbf{R}_i\})\rangle
	e^{-\mathcal{E}_M/ k_B T}, \nonumber \\
    \label{eq:BO}
\end{eqnarray}
where the summation is carried over nuclear quantum states $|\chi_M(\{\mathbf{R}_i\})\rangle$ whose energy is $\mathcal{E}_M$, $\mathcal{Z} = \sum_{M} \exp[-\mathcal{E}_M / k_B T]$ is the partition function of phonons and $k_B$ is the Boltzmann constant. 
In this paper, the observable of interest is the electronic band gap $E_g$ but the formalism applies to any other observables. Similar formulas can also be derived for matrix elements $\mathcal{A}_{jk}(\{\mathbf R_i\})$ between different electronic states; the resultant expectation values will determine the transition rates such as optical absorption and dielectric coefficients~\cite{williams51,lax52}. 


\begin{figure*}
\includegraphics[width=1.95\columnwidth]{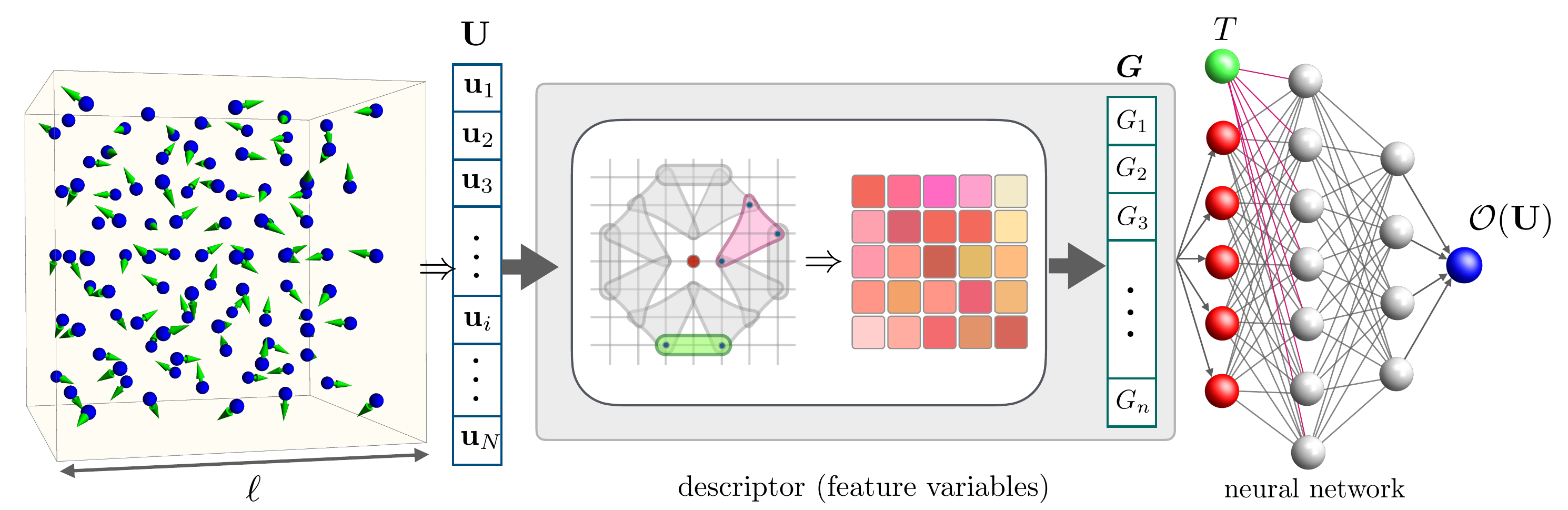}
\caption{Machine learning model for band gap prediction for a given atomic displacement field $\mathbf u_i$ within the a box of linear size~$\ell$. The ML model is composed of two central components: the descriptor and the neural network. The input of the ML model is a cubic block of displacement vectors $\mathbf u_i$. The descriptor corresponds to a representation or feature variables $\bm G = (G_1, G_2, G_3, \cdots)$ of the displacement vectors  that is invariant under symmetry operations of the point group of the lattice, which in the case of Si is the $T_d$ group. The complex dependence of the band gap on the nuclei configurations is encoded in the NN which takes the symmetry-invariant feature variables $\bm G$ as well as the temperature $T$ as the input, and the predicted gap energy at the output. }
\label{fig:nn-model}  \
\end{figure*}


In the harmonic approximation, the quantum number of the nuclei is given by an array of integers $M = \{ n_{\nu, \mathbf k} \}$, and the corresponding wave function becomes a product state: $\chi_M(\{\mathbf R_i\}) \propto \prod_{\nu, \mathbf k} H_{n_{\nu, \mathbf k}}(\xi_{\nu, \mathbf k}) \exp[-\xi^2_{\nu, \mathbf k}/2]$, where $H_n(x)$ is the $n$-th order Hermite polynomial, $\mathbf R_i = \mathbf R^{(0)}_i + \mathbf u_i$, and the displacement vectors $\mathbf u_i$ is expressed in terms of the normal mode coordinates $\xi_{\nu, \mathbf k}$ through Eq.~(\ref{eq:normal_modes}). As the atomic displacements  will play a central role in this work, we introduce the notation $\mathbf U = \{\mathbf u_1, \mathbf u_2, \cdots, \mathbf u_N\}$ to denote the collection of displacement vectors. Within the quadratic approximation, the summation over the integer quantum numbers $n_{\nu, \mathbf k}$ in Eq.~(\ref{eq:BO}) can be recast into integrals over the normal mode coordinates:
\begin{eqnarray}
   \langle \mathcal{O(T)}\rangle = \prod_{\nu, {\bf k}} \int d\xi_{\nu,{\bf k}} \frac{\exp(-\xi_{\nu, \mathbf k}^2/2\sigma^2_{\nu {\bf k},T})}{\sqrt{2\pi}\sigma_{\nu {\bf k},T}} \mathcal{O}({\bf U}),
    \label{eq:gaussian}
\end{eqnarray}
where the width $\sigma_{\nu, \mathbf k}$ of the Gaussian integral is given by:
\begin{eqnarray}
    \sigma^2_{\nu {\bf k},T} = (2n_{\nu {\bf k},T} + 1) \left(\frac{\hbar}{2 \Omega_{\nu,{\bf k}}}\right)^2,
\end{eqnarray}
 and $n_{\nu {\bf k} ,T} = [e^{\hbar \Omega_{\nu {\bf k}}/k_B T}-1]^{-1}$ is the Bose-Einstein occupation of the $(\nu, \mathbf k)$ phonon.  The displacement configuration $\mathbf U$ in $\mathcal{O}(\mathbf U)$ of the Gaussian integrals is again expressed as superposition of the phonon modes as in Eq.~(\ref{eq:normal_modes}).

The computation of the multidimensional Gaussian integral, however, is computationally challenging for observables with a complex dependence on the nuclei configuration. The calculation can be simplified by expanding the observables as a power series expansion of displacement vectors, often up to second order, the Gaussian integral can then be analytically evaluated. Within the DFT framework, the expansion coefficients can be similarly computed using DFPT~\cite{Baroni01,Baroni87,Giannozzi91,Gonze97} or frozen-phonon methods~\cite{Yin80,Fleszar85}.   

An alternative approach, which also allows one to go beyond perturbation theory and DFT, is the stochastic Monte Carlo method~\cite{zacharias15,patrick13,monserrat14}. To this end, we rewrite the integral in Eq.~(\ref{eq:gaussian})  as
\begin{eqnarray}
	\label{eq:MC-average}
	\langle \mathcal{O}(T) \rangle = \int \mathcal{D} \mathbf U \, \pi_G(\mathbf U) \mathcal{O}(\mathbf U) 
	\approx \frac{1}{\mathcal{N}} \sum_{k=1}^{\mathcal{N}} \mathcal{O}({\bf U}^{(k)}).
\end{eqnarray}	
Here $\mathcal{D} \mathbf U$ is a simplified notation for the multiple integrals over the effective coordinates $\xi_{\nu, \mathbf k}$, and $\pi_G(\mathbf U)$ is an effective probability density function of $\mathbf U$, corresponding to the product of Gaussian density functions. The Metropolis-Hastings algorithm is then used to implement a transition probability $P(\mathbf U \to \mathbf U') =  {\rm min}\left[ 1, \pi_G(\mathbf U') / \pi_G(\mathbf U) \right]$ for sampling the displacement configurations. A series of displacement configurations $\mathbf U^{(k)}$ sampled from the Markov-Chain Monte Carlo is used to compute the vibrational average as indicated by the approximation in Eq.~(\ref{eq:MC-average}) above.

\subsection{ML  configuration-to-property models}


Having obtained an {\em ab initio} phonon model, sampling of the nuclei configuration, or more specifically the displacement field $\mathbf u$, is relatively straightforward. The bottleneck of the stochastic approach to vibrational thermal averaging is the computation of the observable $\mathcal{O}(\mathbf U)$ for the sampled configurations. As discussed in Sec.~\ref{sec:intro}, to overcome this computational complexity, a fully deterministic method based on the idea of effective vibrational configurations is proposed to replace the stochastic MC sampling. It has been shown that the vibrational average can be approximated by either a single nuclear configuration, called special displacement configuration~\cite{zacharias16}, or a reduced MC sampling along so-called thermal lines~\cite{monserrat16}. While these methods greatly simplify the problem and have been shown to work reasonably well for a wide range of materials~\cite{Karsai_NJP18}, they are essentially based on quadratic approximation of electron-phonon theory and therefore may fail to capture higher order effects in certain class of materials~\cite{antonius15}. In addition, it has been shown very recently that the special displacement method fails to correctly account for the renormalization of energy levels of nitrogen vacancy centers in diamond~\cite{Kundu24}. 
Hence, it is desirable to develop an efficient vibrational averaging method without sacrificing too much accuracy.

Modern ML methods, especially supervised learning, offer a promising solution to the issue of computational efficiency by providing an accurate and efficient mapping from the phonon configuration $\mathbf U$ to the electronic observable $\mathcal{O}$ of interest. Fundamentally, this approach relies on the universal approximation theorem which shows that deep multilayer neural networks (NN) can be trained to accurately approximate any given high-dimensional functions~\cite{Cybenko89,Hornik89}. A schematic diagram of our proposed ML model is shown in FIG.~\ref{fig:nn-model}; the input of the model is a frozen phonon configuration represented by a displacement field $\mathbf U$ and the temperature $T$, while the output is the predicted observable corresponding to the expectation value $\mathcal{O}(\mathbf U) = \langle \psi | \hat{\mathcal{O}} | \psi \rangle$ where $|\psi \rangle$ is calculated from some electronic structure methods. Using an {\em ab initio} phonon model discussed in Sec.~\ref{sec:ab-initio-phonon} to generate a series of displacement fields $\mathbf U^{(r)}_T$ at a few pre-determined temperatures $T$, the electronic states $|\psi^{(r)}\rangle$ are then solved consistently based on the same first-principle methods. A dataset of the form $\left\{ \mathbf U^{(r)}_T, T; \, \mathcal{O}^{(r)}  \right\}$ is used to train the above ML model.

Broadly speaking, our proposed ML model for $\mathcal{O}(\mathbf U)$ can be viewed as a special case of ML-based modeling of structure-property relationships in materials science~\cite{Jung19,Fan20,Hundi19,zhang_puhan20,Rasul21,xie2018CGCNN,Tian23}. Yet, while the inputs to most conventional ML structure-property model are static disordered or quasi-random structures of the system, a frozen phonon configuration as represented by a displacement field $\mathbf U$ describes an instantaneous structure of nuclei. In this sense, our ML model is also similar to the ML-based interatomic potential models which maps a dynamical atomic configuration in a finite range to a local atomic energy~\cite{behler07,bartok10,li15,behler16,shapeev16,botu17,smith17,zhang18,deringer19}.

%


As shown in Fig.~\ref{fig:nn-model}, there are two central components of the ML model: (i) a learning model based on deep multi-layer neural networks (NN) and (ii) a descriptor which provides generalized or effective coordinates of frozen phonons. The ML model works as follows. First, the set of displacement vectors $\mathbf U = \{\mathbf u_i\}$ that characterizes the atomic configuration within a cubic box is mapped to a collection of symmetry-invariant feature variables $\bm G = \{ G_1, G_2, \cdots \}$, also known as a descriptor. These feature variables are then fed into the NN which produces the predicted observable $\mathcal{O}(\mathbf U)$ at the output node, thus bypassing the time-consuming electronic structure calculations. Integrating this ML phonon-to-property model into the stochastic framework offers an efficient yet accurate vibrational thermal average.

\section{Descriptors for frozen phonon configurations}

\label{sec:descriptor}

\begin{figure*}
\includegraphics[width=1.8\columnwidth]{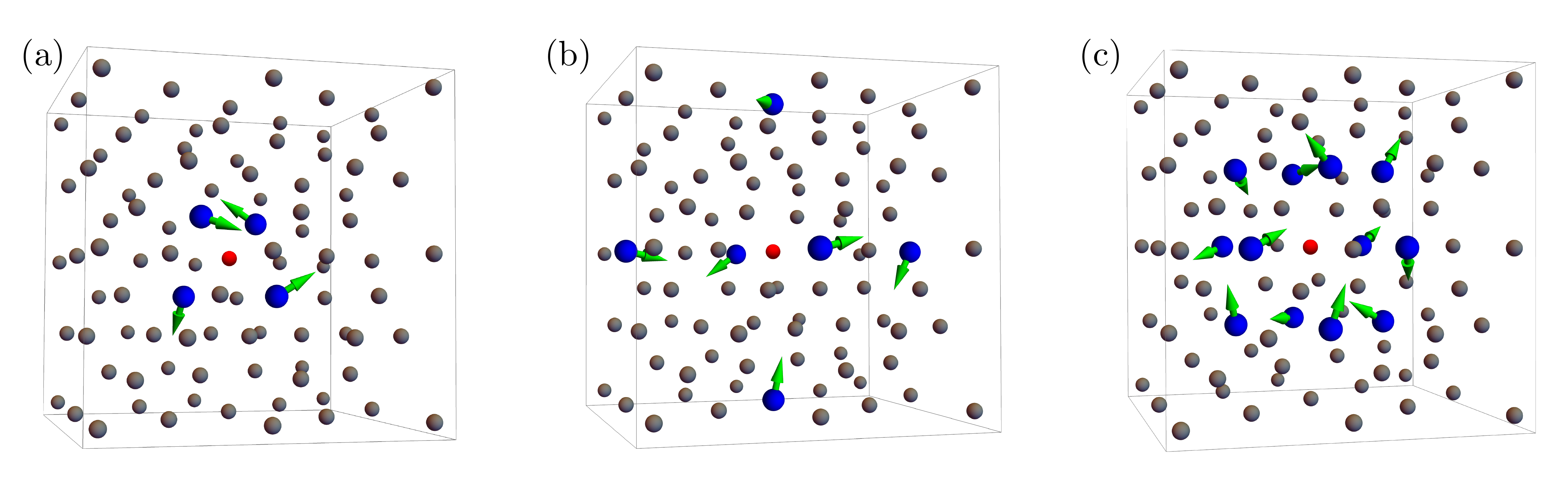}
\caption{Group of atoms sharing same distance to the center of the box form an invariant block of the representation matrix of the symmetry group. Panels (a), (b), and (c) show such invariant groups of 4, 6, and 12 atoms, respectively, in a diamond lattice.}
\label{fig:nb-block}  \
\end{figure*}


While a NN-based learning model is used mainly because of its superb approximation capability, the descriptor plays a crucial role not only to facilitate the training efficiency, but also to preserve the symmetry of the original electron-phonon system. 
In conventional image processing models, the so-called data augmentation technique is commonly employed to artificially expand the size and diversity of a training dataset by applying various transformations to the original images. These transformations can include rotation, scaling, flipping, translation, or color changes, among others. The goal is to make the model more robust and invariant to these transformations, which helps improve generalization and performance on unseen data. However, despite the universal approximation capability of NN models, symmetries of the original physical Hamiltonian can only be learned approximately even with the augmented datasets.

To ensure that the symmetry of the original electronic system is exactly preserved in the ML model, a proper representation of the frozen phonon configuration is to be fed into the learning model instead of the raw displacement vectors. Indeed, similar symmetry-based descriptors also play a crucial role in both ML structure-property models as well as ML force-field models for {\em ab initio} MD simulations. This is particularly important for the BP type ML schemes for interatomic potentials. The output of such BP-type ML models is an atomic energy which, as a scalar, is invariant under symmetry operations of the crystalline system including rotation, reflection, and permutation of same-species atoms. Consequently, a proper descriptor of the local neighborhood not only should be able to differentiate distinct atom configurations, but also remain invariant under the above symmetry transformations.  

Over the past decade, several learning models based on a variety of atomic descriptors have been proposed~\cite{behler07,bartok10,li15,behler11,ghiringhelli15,bartok13,drautz19,himanen20,huo22}. For example, the atom-centered symmetry functions (ACSF), which is one of the most popular atomic descriptors, is based on the two-body (relative distances) and three-body (relative angles) invariants of an atomic configuration~\cite{behler07,behler16}. The group-theoretical method, on the other hand, offers a more controlled approach to the construction of atomic representation based on the bispectrum coefficients~\cite{bartok10,bartok13}. 

In our ML phonon-to-property models, the continuous translation and SO(3) rotation symmetries of crystalline systems are reduced to discrete point group symmetry associated with a supercell. On the other hand, the discrete rotation, reflection, and mirror symmetries of the point group not only transform the atoms in the supercell, but also act uniformly on all displacement vectors $\mathbf u_i$. The output of the ML models is often a scalar, such as band gap energy or optical absorption coefficient. As these quantities are invariant under symmetry operations of the point group, symmetry requirement means that two atomic displacements $\mathbf U^{(a)}$ and $\mathbf U^{(b)}$ related by symmetry operations, when fed into the ML model, should produce exactly the same scalar observable. It is worth noting that, in the absence of descriptors, such symmetry-related configurations will be treated by the neural net as unrelated inputs. To ensure the invariance of the predicted scalar quantity (gap energy), the feature variables $\bm G =  \{ G_1, G_2, \cdots \}$ have to remain invariant under symmetry operators of the point group associated with the lattice.

A general theory of descriptors for lattice systems was recently developed based on group-theoretical methods~\cite{Ma19,zhang22}. Implementations of specific descriptors have also been demonstrated mostly for lattice model systems in condensed matter physics~\cite{Liu22,zhang21,zhang22b,zhang23,cheng23}. Here we adopt the group-theoretical approach to develop a symmetry-invariant descriptor for the frozen phonon configuration in a supercell. To this end, the first step is to decompose a given displacement field $\mathbf U$ into irreducible representations, from which invariant feature variables can be obtained. In the case of silicon where atoms reside on a diamond lattice, the relevant point group of the site symmetry is $T_d$. Take an arbitrary lattice point as the center of reference frame, under a discrete rotation, represented by the $3\times 3$ orthogonal matrix $\mathcal{A}$, the displacement vectors transform according to $\mathbf u_i \, \to \mathbf u'_j = \mathcal{A} \cdot \mathbf u_i$ where the two lattice points are related by $\mathbf R_j = \mathcal{A} \cdot \mathbf R_i$. For $N$ atoms in the box, their displacement field thus constitute a $3N$-dimensional reducible representation of the $T_d$ group. Since the distance is preserved by symmetry operations of the point group, the decomposition of this high-dimensional representation can be simplified as the representation matrix is automatically block-diagonalized, with each block corresponding to a fixed distance from the center-site of the block. FIG.~\ref{fig:nb-block} shows examples of such invariant group of atoms in the diamond lattice. 

Since there are three components for each atomic displacement vector $\mathbf u_j$, the dimension of each block is related to the number of atoms $n_{b}$ in the corresponding neighboring group through the relation $D_b = 3 \times n_b$. For the diamond lattice, there are four distinct types of neighboring groups with $n_b = 4$, 6, 12, and 24 atoms. Consider the simplest case of a 4-atoms group, as shown in FIG.~\ref{fig:nb-block}(a). The displacement vectors of the four atoms, denoted as $\mathbf a$, $\mathbf b$, $\mathbf c$, and $\mathbf d$ for simplicity, form a  $D_b = 12$ dimensional representation of the $T_d$ group, which can be decomposed as $12=A_1\bigoplus E\bigoplus T_1\bigoplus 2T_2$, where, $A_1$ denotes 1D representation, $E$ denotes 2D representation and $T$ denotes 3D representation. The $A_1$ component, for example, is given by the combination
\begin{eqnarray*}
	& & f^{A_1} =a_x+a_y+a_z-b_x-b_y+b_z \nonumber \\
	& & \qquad +c_x-c_y-c_z-d_x+d_y-d_z.
\end{eqnarray*}
Another example is the basis functions of one of the $T_2$ IR:
\begin{eqnarray*}
	& & f^{T_2}_x = a_x+b_x+c_x+d_x, \\
	& & f^{T_2}_y = a_y+b_y+c_y+d_y, \\
	& & f^{T_2}_z = a_z+b_z+c_z+d_z. 
\end{eqnarray*}
Details of the IR decompositions and the basis functions can be found in Appendix~\ref{app:descriptor}.

By repeating the same procedures for each block, the displacement field $\mathbf U$ is fully decomposed into the IR of the $T_d$ group. For convenience, we arrange the resultant IR basis functions into a vector 
\begin{eqnarray}
	\bm{f}^\Gamma_r=(f^\Gamma_{r,1},f^\Gamma_{r,2},\cdots,f^\Gamma_{r,D_{\Gamma}}), 
\end{eqnarray}
where $\Gamma$ labels the type of IR, $r$ enumerates the multiple occurrence of IR-$\Gamma$ in the decomposition of the displacement configuration, and $D_{\Gamma}$ is the dimension of the corresponding IR. Given these basis functions, one can immediately obtain a set of invariants called power spectrum $\{p^\Gamma_r\}$, which are the amplitude of each individual IR basis, i.e. 
\begin{eqnarray}
	p^\Gamma_r = \left| \bm f^\Gamma_r  \right|^2. 
\end{eqnarray}
The power spectrum coefficients constitute a basic set of feature variables that are invariant under symmetry transformations of the point group. However, a descriptor based only on the power spectrum is incomplete in the sense that the relative phases between different IRs are ignored. For example, the relative ``angle" between two IRs of the same type: $\cos\theta = (\bm{f}^{\Gamma}_{r_1}\cdot\bm{f}^{\Gamma}_{r_2})/|\bm{f}^{\Gamma}_{r_1}||\bm{f}^{\Gamma}_{r_2}|$ is also an invariant of the symmetry group. Without such phase information, the NN model might suffer from additional error due to the spurious symmetry, namely two IRs can freely rotate independent of each other.

A systematic approach to include all relevant invariants, including both amplitudes and relative phases, is the bispectrum method~\cite{kondor07,bartok13}. The bispectrum coefficients are triple product of IR basis functions defined as
\begin{eqnarray}
	\label{eq:bispectrum}
	B^{\Gamma, \Gamma_1, \Gamma_2}_{r, r_1, r_2} = C^{\Gamma; \Gamma_1, \Gamma_2}_{\alpha, \beta,\gamma} f^{\Gamma}_{r, \alpha} f^{\Gamma_1}_{r_1, \beta} f^{\Gamma_2}_{r_2, \gamma}
\end{eqnarray}
where $C^{\Gamma; \Gamma_1, \Gamma_2}$ are the Clebsch-Gordan coefficients of the point group~\cite{kondor07}. 
While a bispectrum descriptor provide a faithful invariant description of the phonon field, the number of all bispectrum coefficients increases cubically with the size of supercell. Moreover, many of them are redundant.  In this work, we develop a descriptor similar to the one used in Ref.~\cite{Ma19,zhang22}, that is modified from the bispectrum method. We introduce a set of reference basis functions $\bm f^{\Gamma}_{\rm ref}$ for each IR $\Gamma$ of the point group. These reference basis are computed by averaging large blocks of displacement variables, such that they are less sensitive to small changes in the supercell. For example, by dividing the supercell into 24 symmetry-related blocks for the case of $T_d$ point group, we define the average displacement vector of a block-$B$ as $\overline{\mathbf u}_B = \sum_{i \in B} \mathbf u_i$. The reference basis $\bm f^\Gamma_{\rm ref}$ can then be computed from these block-averaged displacement vectors using exactly the same decomposition formulas. 

Given the reference IR, we then define the relative ``phase" $\theta^\Gamma_r$ of an IR as the projection of its basis functions onto the reference basis: 
\begin{eqnarray}
	\exp\left( i\theta^\Gamma_r \right) \equiv \bm f^\Gamma_r \cdot \bm f^\Gamma_{\rm ref} / |\bm f^\Gamma_r |\, |\bm f^\Gamma_{\rm ref}|. 
\end{eqnarray}	
The relative phases between two IR basis of the same type can then be obtained via their respective phases with the reference. Finally, the relative phases between {\em different} IR basis are provided by the bispectrum coefficients computed entirely from the reference IRs. To summarize, the lattice descriptor are comprised of the following three types of variables: 
\begin{eqnarray}
	\bm G = \{ G_\ell \} = \{ p^{\Gamma}_r , \, e^{i \theta^\Gamma_r}, B^{\Gamma, \Gamma_1, \Gamma_2}_{\rm ref} \}. 
\end{eqnarray}
Here $B^{\Gamma, \Gamma_1, \Gamma_2}_{\rm ref}$ is the bispectrum coefficient in Eq.~(\ref{eq:bispectrum}) where all three IRs are computed from the reference. 
These feature variables are not only invariant under the point-group symmetry, but also provide a faithful representation of phonon configurations (module symmetry operations) in the supercell.

\section{Computational details and results}

\label{sec:results}

\subsection{DFT Calculations}
First principles density-functional-theory (DFT) calculations were done using Quantum Espresso (QE)~\cite{QE-2009} package. 
Perdew-Zunger (PZ) exchange-correlation functional~\cite{Perdew81} within the local density approximation (LDA) was used in all the calculations.
Full lattice relaxation of primitive cell of Si crystal was performed which gave lattice parameter of 5.398 \AA~consistent with previous works~\cite{Monserrat18}.
The primitive BZ was sampled by using \textit{k}-mesh of 24 $\times$ 24 $\times$ 24 and energy cutoff of 30 Rydberg was used after careful convergence tests. 
The relaxed lattice was used to perform {\em ab initio} phonon calculation using DFPT~\cite{Baroni01,Giustino07,Bernardi16} within the harmonic approximation as implemented in QE. 
For phonon calculations, we used a \textit{q}-mesh of 8 $\times$ 8 $\times$ 8.
From the information of phonon eigen-frequencies $\Omega_{\nu, \mathbf k}$ and eigenvectors $\mathbf V^{(\nu)}_{s, \mathbf k}$, we generated atomic supercell configurations of size 6 $\times$ 6 $\times$ 6 primitive unit cells containing 432 Si atoms using importance sampling Monte Carlo according to Eq.~(\ref{eq:gaussian}). 
Subsequently, DFT self-consistent calculation were performed on about 10\% of the distorted atomic supercell configurations using \textit{k}-mesh of 3 $\times$ 3 $\times$ 3 to generate a database of electronic band gap as a function of atomic configurations.



\begin{figure}[t]
            \includegraphics[width=0.99\columnwidth]{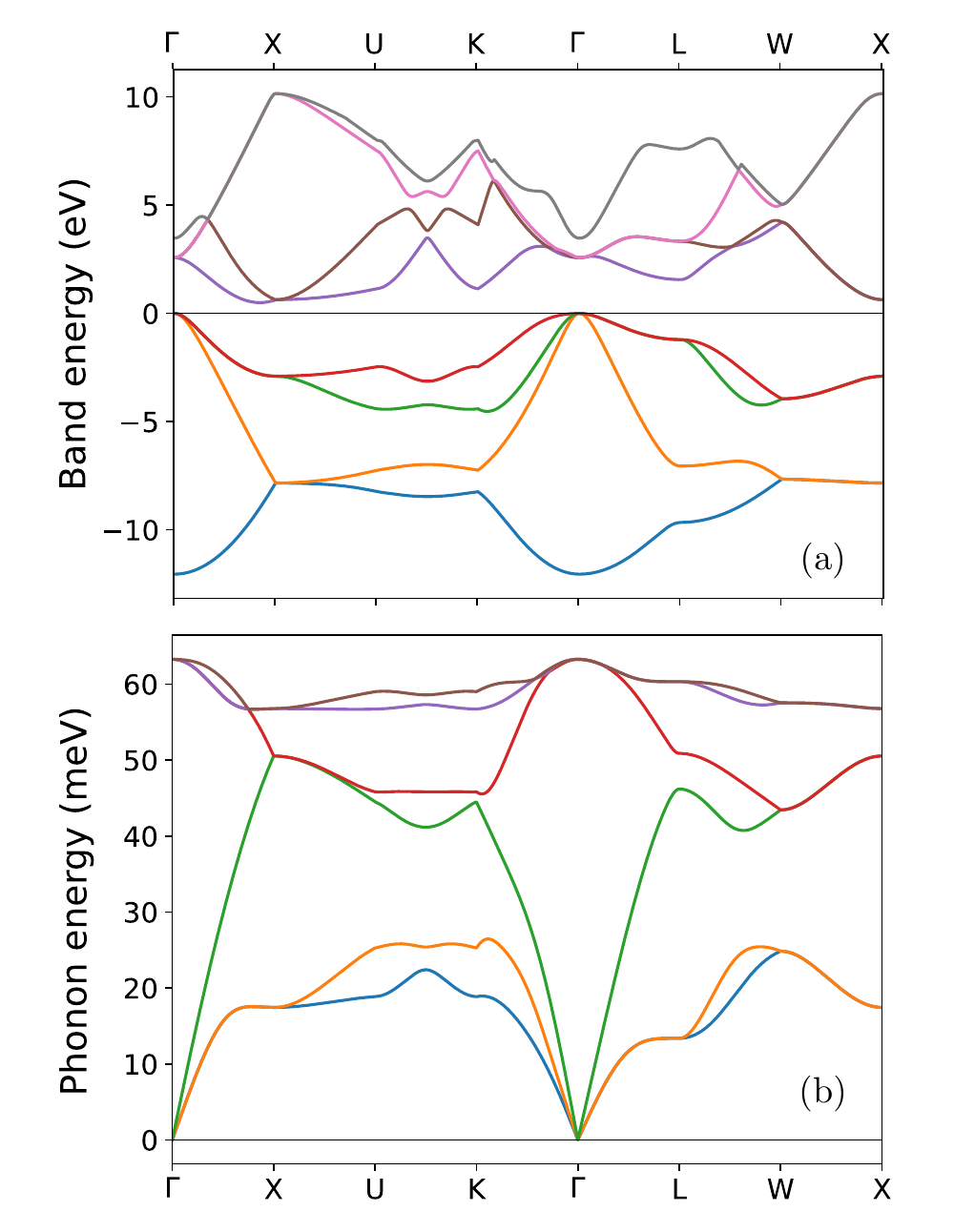}
            \caption{Ground state (a) electron and (b) phonon dispersion of Si crystal along high-symmetry directions of the lattice.
        }
        \label{fig:dispersion} 
\end{figure}

In FIG.~\ref{fig:dispersion}, we show the electron and phonon dispersion relations in the ground state of silicon crystal obtained from DFT calculations. 
These results are consistent with previous literature~\cite{Si_phonon_Wei_PRB1994}. 
DFT calculations show an indirect band gap of 500 meV which is severely underestimated compared to the experimental band gap of 1.12 eV. Such band gap underestimation of insulators and semiconductors is a well known problem in DFT and more sophisticated methods like GW calculations are known to give better estimation of the band gap. In practice, GW correction of band gap is similar to the ``scissor shift'' of bands. 
Since we are interested in the band gap reduction induced by thermal phonons, this underestimation of band gap should not be an issue as the same GW-correction (to leading order) appears in both the perfect and perturbed crystal structures.
Hence, in the following, we will use phonon induced band gap correction $\Delta E_g=E_g^{\rm eq} - E_g^{\rm distorted}$  as our observable of interest, where $E_g^{\rm eq}$ is the band gap of the perfect crystalline structure and 
$E_g^{\rm distorted}$ is the band gap of the distorted structure.

\begin{figure}[t]
            \includegraphics[width=0.48\textwidth]{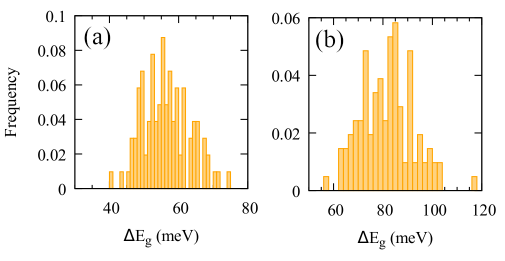}
            \caption{Histogram showing distribution of phonon-induced band gap correction at (a) $T = 0 $K and (b) $T = 200$ K using 100 data points for Si-supercell of size 6 $\times$ 6 $\times$ 6.
        }
        \label{fig:histogram} 
\end{figure}

The phonon dispersion shown in FIG.~\ref{fig:dispersion} is also consistent with previous studies. 
By sampling the normal mode coordinates $\xi_{\nu \mathbf k}$ according to Eq.~(\ref{eq:gaussian}), a displacement configuration $\mathbf U = \{ \mathbf u_i \}$, or a frozen phonon configuration, within the supercell is constructed according to Eq.~(\ref{eq:normal_modes}). Fixing the nuclei positions at $\mathbf R_i = \mathbf R_i^{(0)} + \mathbf u_i$ in the supercell, DFT calculations were performed to obtain the optimized electron density distribution. An energy gap $E_g(\mathbf U)$ corresponding to a particular phonon configuration is computed as the difference between the lowest unoccupied and the highest occupied states.
In FIG.~\ref{fig:histogram}, we show histogram plot of the band gap correction $\Delta E_g$ for a sample size of 100 at two different temperature values. 
As seen from the plots, the MC method has not fully converged and needs large number of samples to get a better estimation of the band gap.
In the following, we show that it is possible to get a better estimation (i.e. smaller standard deviation) of the band gap  without incurring additional computational cost by first using the DFT calculated dataset to train a ML phonon-to-band gap mapping, and then using the ML model to predict band gap for a large number of configurations to obtain thermal average.

\subsection{Neural network architecture and model training}

\label{sec:nn}

As shown in FIG.~\ref{fig:nn-model}, our learning model is based on a fully-connected neural network (NN), also known as a multilayer perceptron model. A fully-connected NN comprises of a series of fully connected linear layers with non-linear activation functions that are applied at each layer. The parameters of the $j$-th layer are the matrix of network weights $\mathbf W^{(j)}$ and bias vector $\mathbf b^{(j)}$. The weights introduce coupling between neurons of adjacent layers. The mapping of neurons $\bm x$ from one layer to the next is achieved by a linear transformation followed by a nonlinear activation function as
\begin{eqnarray}
	x_m^{(j+1)} = f_{\rm av}\!\left( \sum_{n} W^{(j)}_{mn} \, x^{(j)}_n + b^{(j)}_m \right). 
\end{eqnarray}
A commonly used activation function is the rectified linear unit (ReLU)~\cite{Agarap19} defined as $f_{\rm av}(x) := \max(x,0)$. However, sometimes a large number of ReLU neurons remains inactivated in the model. Hence, often some variants of leaky ReLU~\cite{Maas13} are used instead. In this work, we used the Gaussian Error Linear Units (GeLU)~\cite{Hendrycks23} activation function, which shows better performance than the standard ReLU. 

\begin{table}[t]
\label{tab:table1}
\begin{center}
\begin{ruledtabular}
\begin{tabular}{c|cc}
\textrm{Layer}&\textrm{Network}\\
\colrule
Input Layer & [838,2048]\footnote{Fully connected layer with arguments [input size, output size].} \\
Hidden Layer 1 & [2048,1024] \\
Hidden Layer 2 & [1024,512] \\
Hidden Layer 3 & [512,256] \\
Output Layer & [256,1]  \\ \hline
\end{tabular}
\end{ruledtabular}
\caption{The architecture of neural network. GeLU activation function and drop-out rate of 0.3 was used in all layers except output layer.}
\end{center}
\end{table}

We build a fully-connected NN with four hidden layers. The dimension of the hidden layers are (2048, 1024, 512, 256); see Table~I for details. Moreover, a dropout layer with 0.3 dropout rate is introduced. As for the initialization of each layers, we utilize the Xavier uniform for the weights and the Normal distribution for the bias. Adam optimizer with initial 0.0001 cosine learning rate and $l_2$ regularization coefficient $5\times10^{-9}$ is used.

As noted above, the phonon configurations are  sampled from a supercell consisting of $6 \times 6 \times 6$ primitive unit cells. Here a primitive unit cell corresponds to a parallelepiped formed from primitive vectors $\mathbf a_1 = (0, a/2, a/2)$, $\mathbf a_2 = (a/2, 0, a/2)$, and $\mathbf a_3 = (a/2, a/2, 0)$, where $a$ is the length of a conventional cubic unit cell. The skewed primitive unit cell, however, does not preserve the cubic symmetry of the silicon crystal. To facilitate the incorporation of the point-group symmetry, here we instead use an input block comprised of $3 \times 3 \times 3$ cubic cells with a total of 279 atoms.

To further improve the statistics of the ML predictions, we apply the same ML model to multiple overlapping cubic blocks of the supercell. Specifically, we choose sites with coordinate $(i, j, k) a$ and $(i+1/2, j+1/2, k+1/2) a$ as the center, where $i, j, k$ are integers ranging from 0 to 2. For these 54 sites, we consider their surrounding sites within a block of 3 $\times$ 3 $\times$ 3 cubic cells which includes 279 sites. Since our descriptors keep the same dimension as the configuration,  the number of the input features is $838 = 279\times 3 +1$, where first 837 features are the descriptors of 279 sites configuration and the rest feature is the temperature. The band gap energy corresponding to the phonon configuration $\mathbf U$ of the supercell is obtained by averaging ML predictions from these 54 blocks. Accordingly, the loss function used for training the NN is given by the following mean square error (MSE)
\begin{equation}
    L= \Biggl\langle \Biggl| E^{\rm DFT}_{g} - \frac{1}{M} \sum_{k=1}^M \hat{E}_g^{(k)} \Biggr|^2 \Biggr\rangle,
\end{equation}
where $\hat{E}_g^{(k)}$ denotes the ML predicted band gap energy for the $k$-th cubic block, $M = 54$ is the number of blocks, and $\langle \cdots \rangle$ indicates averaging over the training dataset. We have generated 103, 108, 113, and 204 distinct phonon configurations for four different temperatures at $T=0, 100, 200$, and 300K, respectively, and performed fully self-consistent DFT calculations on these configurations to estimate the band gap $E^{\rm DFT}_g$. For each temperature, 80 randomly chosen configurations and the corresponding band gap energies are used to train the ML model, while the rest of the configurations are kept as the validation and test dataset. After 600 epochs of training process with each batch including only one configuration, the loss function value is converged to around 0.0024297 for the training dataset and 0.00410986 for the validation dataset. The whole process takes 540 seconds on  NVIDIA GPU~A100 workstation.


\begin{figure}[t]
\includegraphics[width=0.75\columnwidth]{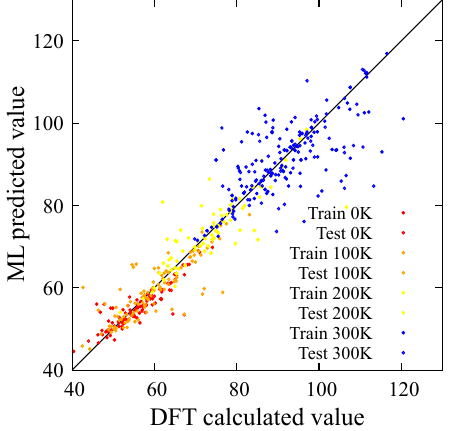}
\caption{Scatter plot showing phonon-induced band gap correction (in meV units) predicted by ML model versus DFT calculation for $6\times 6 \times 6$ Si-supercell at four different temperature values.}
\label{fig:scatter-plot}
\end{figure}

\begin{figure}[htb]
\includegraphics[width=0.99\columnwidth]{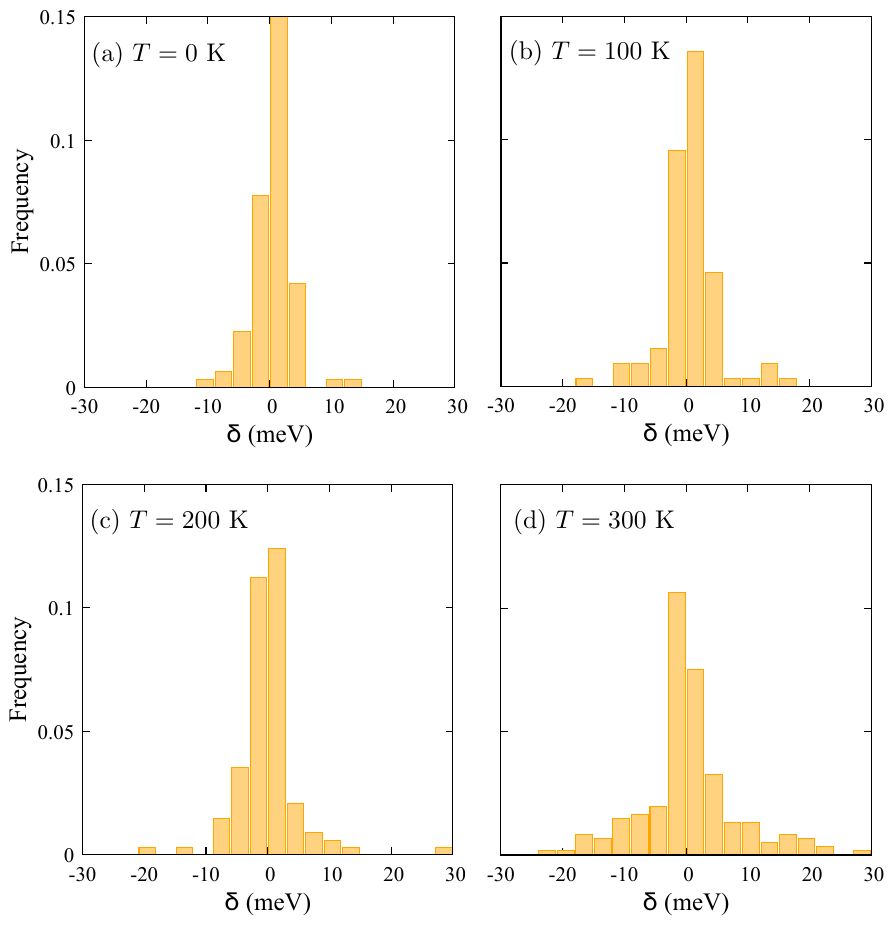}
\caption{Histogram of the prediction error defined as $\delta = \Delta E_g^{\rm ML} - \Delta E_g^{\rm DFT}$ for temperatures at (a) $T = 0$~K, (b) 100~K, (c) 200~K, and (d) 300~K.}
\label{fig:hist-error}
\end{figure}

FIG.~\ref{fig:scatter-plot} shows the parity plot of ML predicted band gap corrections $\Delta_g$  versus those from DFT calculations for four temperatures used in NN training. As expected, the phonon-induced correction to the band gap energy is enhanced at higher temperatures due to the increased number of thermal phonons. While an overall agreement was obtained between the ML predicted values and DFT results, the prediction accuracy varies with the temperatures. For a more detailed comparison, histogram pots of the prediction error $\delta = \Delta E_g^{\rm ML} - \Delta E_g^{\rm DFT}$ are shown in FIG.~\ref{fig:hist-error}(a)--(d) for four different temperatures. The error is larger for higher temperature because the width of the Gaussian distribution used in MC sampling increases with temperature. Yet, it is remarkable that the error from NN trained on $\sim$ 100 data points is merely of the order of 10~meV, thus demonstrating the robustness of our ML model.


\begin{figure}[t]
        \includegraphics[width=0.87\columnwidth]{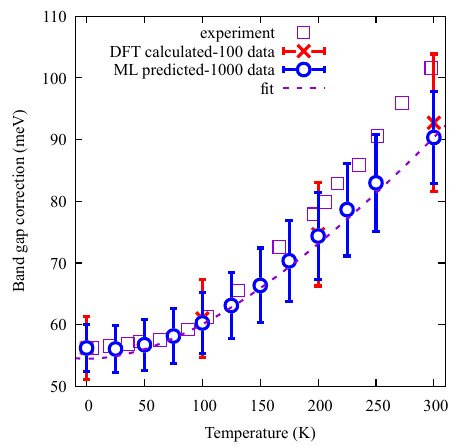}
        \caption{ Temperature dependence of the phonon-induced band-gap correction $\Delta E_g$ of Si. In total, 528 DFT calculations on 432-atom Si supercell were done. ML prediction was done on 1000 configurations at each temperature. Experimental data is extracted from Ref.~\cite{Alex96}. The experimental data is shifted by DFT calculated zero-point band gap correction energy.}
    \label{fig:T_bandgap} 
\end{figure}

After NN training and optimization, we used the trained NN to predict band gap for a large number of MC generated supercell configurations at the four temperature values at which DFT calculations were performed to train the ML model.  FIG.~\ref{fig:T_bandgap} shows the temperature dependence of the phonon-induced band gap correction of Si using DFT and NN prediction. For comparison, we also show the experimentally measured band gap correction. It should be noted that, although there are noticeable discrepancies between experiment and computational results for higher temperature region ($T \gtrsim 150$ K), the ML predictions agree quite well with the DFT calculations. Importantly, by using ML-predicted data set, we were able to reduce the errorbars at $T = 0, 100, 200$ and 300~K significantly compared to the DFT calculation.

It is remarkable that both DFT calculation and ML prediction correctly capture the temperature dependence of band gap, especially in the low temperature regime. The discrepancy at higher temperature is likely due to the anharmonic and lattice thermal expansion effects which are not incorporated in our calculations. Inclusion of these effects in \textit{ab initio} calculations and application of more accurate exchange correlation functionals could yield better comparison with experiments\cite{SCP_Souvatzi2009,anharmonic_Zacharias2020}. As discussed above, by training a ML model using more accurate {\em ab initio} data, our proposed framework can also incorporate such many-body effects into the electron-phonon couplings and the phonon-induced temperature dependence.

Moreover, since temperature serving as a conditional control is another input to our NN (see FIG.~\ref{fig:nn-model}), we also apply the trained ML model to predict the band gaps at intermediate temperature values (see FIG.~\ref{fig:T_bandgap}). The inclusion of temperature as a control parameter to the ML model, which was trained by datasets from multiple temperatures, could impose further constraints by enforcing a consistency between predictions at different temperatures. We have also trained ML models only trained by dataset from a particular temperature and obtained better prediction accuracies at that temperature. However, such models not only lack the generality for other temperatures, but also likely suffer from over-fitting. On the other hand, the ML models with additional conditioning from the temperature is shown to produce consistent trend of temperature dependent band gap energy. The prediction accuracies at the intermediate temperatures are of the same order as those used for the training. We note that the overall prediction accuracy can be improved by increasing the number of training data points or the scale of the neural network. Our results show that, by leveraging the power of transfer learning, band gap corrections at intermediate temperatures (where training data are not available) can also be accurately predicted by our trained ML model without incurring any significant computational cost. The only cost is in generating supercell configurations which is nominal.

\section{Conclusion and Outlook}

\label{sec:conclusion}

In summary, we have proposed a ML-based {\em ab inito} framework to incorporate phonon-induced renormalization of electronic band structures and other electronic properties. Our ML framework is based on the well-established stochastic approach to temperature dependent band structures and transition processes due to electron-phonon coupling. In this approach, frozen phonon configurations within a supercell are first sampled from Monte Carlo or dynamical simulations based on an {\em ab initio} phonon model. Supervised learning methods are then applied to build a ML model that accurately maps the sampled frozen phonon configuration within a supercell to the physical properties of interest. With the efficient sampling of phonon configurations and prediction of the corresponding physical properties, accurate vibrational temperature dependence can be obtained based on moderate datasets of {\em ab initio} calculations. 

A crucial component of our ML framework is a phonon descriptor for incorporating the point-group symmetry of the crystalline system into the ML model. Contrary to atomic descriptors used in most ML-based {\em ab initio} molecular dynamics methods, where the objects of interest are the atomic species and relative coordinates to a central atom,  the dynamical variables in our case are atomic displacement vectors (from their equilibrium position in a perfect crystal). To properly account for these differences, we have employed the group-theoretical methods to obtain feature variables for the input to the neural network which is used as the learning model in our framework. 


In this work, we apply the proposed ML framework to obtain the temperature dependence of phonon-induced corrections to the electronic band gap in crystalline silicon. We demonstrate that the prediction errors of the energy gaps can be significantly reduced from less than 100 datasets thereby reducing the number of expensive DFT computations necessary for better estimation. Moreover, applying the trained ML model at temperatures not included in the training dataset, we show that accurate and consistent trend of the band-gap corrections are obtained, highlighting the power of transfer learning. While our work focuses on the prediction of band gap energy, the proposed ML framework is of general purpose and can be used for thermal vibrational effects on other electronic properties or transition coefficients as long as Born-Oppenheimer approximation is valid. 



While the ML model developed in this work is based on a specific flavor of DFT (local density approximation for electron and phonon calculations) and  density-functional perturbation theory-based \textit{ab-initio} phonon model, as discussed above, the frozen phonon approach  is independent of the underlying electronic structure methods. This  means that many-body effects beyond DFT can be readily included in this approach. However, the {\em ab initio} phonon model used to generate the atomic configurations should be consistent with the first-principle method employed for training the ML model.

\begin{acknowledgments}
NA and WY are supported by U.S. Department of Energy (DOE) the Office of Basic Energy Sciences, Materials Sciences and Engineering Division under Contract No. DE-SC0012704. SZ and GWC acknowledge support of the US Department of Energy Basic Energy Sciences under Contract No. DE-SC0020330, as well as the support of Research Computing at the University of Virginia.
\end{acknowledgments}

\bibliography{ref}

\appendix

\begin{widetext} 

\newpage

\section{Details of the Descriptor for displacements on a diamond lattice}
\label{app:descriptor}

The silicon atoms around a centered atom form into four different cases: four points group, six points group, twelve points group, and twenty-four points group. Only former three cases occurred in our case of $2\times 2\times 2$-supercell. Four points case displacement vectors (corner points of the cube) can be decomposed into $4\times3=A_1\bigoplus E\bigoplus T_1\bigoplus 2T_2$ with the following basis:
\begin{align*}
f^{A_1} &=a_x+a_y+a_z-b_x-b_y+b_z+c_x-c_y-c_z-d_x+d_y-d_z \\
f^{E}_x &= a_x+a_y-2a_z-b_x-b_y-2b_z+c_x-c_y+2c_z-d_x+d_y+2d_z \\
f^{E}_y &= \sqrt{3}(-a_x+a_y+b_x-b_y-c_x-c_y+d_x+d_y) \\
f^{T_1}_x &= a_y-a_z+b_y+b_z-c_y+c_z-d_y-d_z \\
f^{T_1}_y &= a_x-a_z+b_x+b_z-c_x-c_z-d_x+d_z \\
f^{T_1}_z &= -a_x+a_y+b_x-b_y+c_x+c_y-d_x-d_y \\
f^{T_2,1}_x &= a_x+b_x+c_x+d_x \\
f^{T_2,1}_y &= a_y+b_y+c_y+d_y \\
f^{T_2,1}_z &= a_z+b_z+c_z+d_z \\
f^{T_2,2}_x &= a_y+a_z+b_y-b_z-c_y-c_z-d_y+d_z \\
f^{T_2,2}_y &= a_x+a_z+b_x-b_z-c_x+c_z-d_x-d_z \\
f^{T_2,2}_z &= a_x+a_y-b_x-b_y-c_x+c_y+d_x-d_y
\end{align*}
where the Cartesian coordinates of those points are $a(0.5, 0.5, 0.5)$, $b(-0.5,-0.5,0.5)$, $c(0.5,-0.5,-0.5)$, and $d(-0.5,0.5,-0.5)$, or $\{-a,-b,-c,-d\}$. 
The six points case displacement vectors (face-centered points of the cube) can be decomposed into $6\times3=A_1\bigoplus E\bigoplus 2T_1\bigoplus 3T_2$ with the following basis:
\begin{align*}
f^{A_1} &=a_x+b_y+c_z-d_z-e_y-f_x \\
f^{E}_x &= a_x+b_y-2c_z+2d_z-e_y-f_x\\
f^{E}_y &= \sqrt{3}(-a_x+b_y-e_y+f_x) \\
f^{T_1,1}_x &= b_x-c_x-d_x+e_x \\
f^{T_1,1}_y &= a_y-c_y-d_y+f_y \\
f^{T_1,1}_z &= -a_z+b_z+e_z-f_z \\
f^{T_1,2}_x &= -b_z+c_y-d_y+e_z \\
f^{T_1,2}_y &= -a_z+c_x-d_x+f_z \\
f^{T_1,2}_z &= a_y-b_x+e_x-f_y \\
f^{T_2,1}_x &= a_x+f_x \\
f^{T_2,1}_y &= b_y+e_y \\
f^{T_2,1}_z &= c_z+d_z \\
f^{T_2,2}_x &= b_z +c_y-d_y-e_z\\
f^{T_2,2}_y &= a_z+c_x-d_x-f_z \\
f^{T_2,2}_z &= a_y+b_x-e_x-f_y \\
f^{T_2,3}_x &= b_x+c_x+d_x+e_x \\
f^{T_2,3}_y &= a_y+c_y+d_y+f_y \\
f^{T_2,3}_z &= a_z+b_z+e_z+f_z \\
\end{align*}
where the Cartesian coordinates of those points are $a(2,0,0)$, $b(0,2,0)$, $c(0,0,2)$, $d(0,0,-2)$, $e(0,-2,0)$, and $f(-2,0,0)$. 
The twelve points case displacement vectors (edge-centered points of the cube) can be decomposed into $12\times3=2A_1\bigoplus A_2\bigoplus 3E\bigoplus 4T_1\bigoplus 5T_2$ with the following basis:
\begin{align*}
f^{A_1,1} &= a_x+a_y+b_y+b_z+c_x+c_z+d_y-d_z+e_x-e_z-f_y+f_z \\
&-g_x-g_z-h_y-h_z-i_x+i_z+j_x-j_y-k_x+k_y-l_x-l_y\\
f^{A_1,2} &= a_z+b_x+c_y-d_x-e_y-f_x+g_y+h_x-i_y-j_z-k_z+l_z\\
f^{A_2} &= a_x-a_y+b_y-b_z-c_x+c_z+d_y+d_z-e_x-e_z-f_y-f_z\\ 
&+g_x-g_z-h_y+h_z+i_x+i_z+j_x+j_y-k_x-k_y-l_x+l_y\\
f^{E,1}_x &= 2a_x+2a_y-b_y-b_z-c_x-c_z-d_y+d_z-e_x+e_z+f_y-f_z \\ 
&+g_x+g_z+h_y+h_z+i_x-i_z+2j_x-2j_y-2k_x+2k_y-2l_x-2l_y \\
f^{E,1}_y &= \sqrt{3}(b_y+b_z-c_x-c_z+d_y-d_z-e_x+e_z\\
&\ \ \ \ \ -f_y+f_z+g_x+g_z-h_y-h_z+i_x-i_z) \\
f^{E,2}_x &= -b_y+b_z-c_x+c_z-d_y-d_z-e_x-e_z\\
&\ \ +f_y+f_z+g_x-g_z+h_y-h_z+i_x+i_z\\
f^{E,2}_y &= \frac{1}{\sqrt{3}}(2a_x-2a_y-b_y+b_z+c_x-c_z-d_y-d_z+e_x+e_z+f_y+f_z\\
&\ \ -g_x+g_z+h_y-h_z-i_x-i_z+2j_x+2j_y-2k_x-2k_y-2l_x+2l_y)\\
f^{E,3}_x &= 2a_z-b_x-c_y+d_x+e_y+f_x\\
&-g_y-h_x+i_y-2j_z-2k_z+2l_z\\
f^{E,3}_y &= \sqrt{3}(b_x-c_y-d_x+e_y-f_x-g_y+h_x+i_y)\\
f^{T_1,1}_x &= a_x-c_x-e_x-g_x-i_x+j_x+k_x+l_x\\
f^{T_1,1}_y &= a_y-b_y-d_y-f_y-h_y+j_y+k_y+l_y \\
f^{T_1,1}_z &= b_z-c_z+d_z-e_z+f_z-g_z+h_z-i_z \\
f^{T_1,2}_x &= a_y-c_z+e_z-g_z+i_z-j_y-k_y+l_y \\
f^{T_1,2}_y &= a_x-b_z+d_z+f_z-h_z-j_x-k_x+l_x \\
f^{T_1,2}_z &= b_y-c_x-d_y+e_x-f_y-g_x+h_y+i_x \\
f^{T_1,3}_x &= -b_y+b_z+d_y+d_z-f_y-f_z+h_y-h_z \\
f^{T_1,3}_y &= -c_x+c_z+e_x+e_z+g_x-g_z-i_x-i_z \\
f^{T_1,3}_z &= a_x-a_y-j_x-j_y+k_x+k_y-l_x+l_y\\
f^{T_1,4}_x &= a_z-c_y+e_y+g_y-i_y-j_z+k_z-l_z \\
f^{T_1,4}_y &= a_z-b_x+d_x-f_x+h_x+j_z-k_z-l_z \\
f^{T_1,4}_z &= b_x-c_y+d_x-e_y-f_x+g_y-h_x+i_y\\
f^{T_2,1}_x &= a_x+c_x+e_x+g_x+i_x+j_x+k_x+l_x\\
f^{T_2,1}_y &= a_y+b_y+d_y+f_y+h_y+j_y+k_y+l_y \\
f^{T_2,1}_z &= b_z+c_z+d_z+e_z+f_z+g_z+h_z+i_z \\
f^{T_2,2}_x &= a_y+c_z-e_z+g_z-i_z-j_y-k_y+l_y\\
f^{T_2,2}_y &= a_x+b_z-d_z-f_z+h_z-j_x-k_x+l_x \\
f^{T_2,2}_z &= b_y+c_x-d_y-e_x-f_y+g_x+h_y-i_x\\
f^{T_2,3}_x &= b_y+b_z-d_y+d_z+f_y-f_z-h_y-h_z \\
f^{T_2,3}_y &= c_x+c_z-e_x+e_z-g_x-g_z+i_x-i_z \\
f^{T_2,3}_z &= a_x+a_y-j_x+j_y+k_x-k_y-l_x-l_y \\
f^{T_2,4}_x &= a_z+c_y-e_y-g_y+i_y-j_z+k_z-l_z \\
f^{T_2,4}_y &= a_z+b_x-d_x+f_x-h_x+j_z-k_z-l_z \\
f^{T_2,4}_z &= b_x+c_y+d_x+e_y-f_x-g_y-h_x-i_y\\
f^{T_2,5}_x &= b_x+d_x+f_x+h_x \\
f^{T_2,5}_y &= c_y+e_y+g_y+i_y \\
f^{T_2,5}_z &= a_z+j_z+k_z+l_z
\end{align*}
where the Cartesian coordinates of those points are $a(1,1,0)$, $b(0,1,1)$, $c(1,0,1)$, $d(0,1,-1)$, $e(1,0,-1)$, $f(0,-1,1)$, $g(-1,0,-1)$, $h(0,-1,-1)$, $i(-1,0,1)$, $j(1,-1,0)$, $k(-1,1,0)$, and $l(-1,-1,0)$.

The feature variables which are invariant under symmetry operations of the point-group $T_d$ are given by the following invariants:
\begin{align*}
G^{A_1} &= f^{A_1} \\
G^{A_2} &= f^{A_2}f^{A_2}_\text{ref}\\
G^{E}_1 &= \langle\boldsymbol{f}^{E},\boldsymbol{f}^{E}\rangle \\
G^{E}_2 &= \langle\boldsymbol{f}^{E},\boldsymbol{f}^{E}_\text{ref}\rangle \\
G^{T_1}_1 &= \langle\boldsymbol{f}^{T_1},\boldsymbol{f}^{T_1}\rangle \\
G^{T_1}_2 &= \langle\boldsymbol{f}^{T_1},\boldsymbol{f}^{T_1}_\text{ref}\rangle \\
G^{T_1}_3 &= f^{A_2}_\text{ref}(f^{T_1}_xf^{T_1}_{\text{ref},y}f^{T_1}_{\text{ref},z} + f^{T_1}_yf^{T_1}_{\text{ref},x}f^{T_1}_{\text{ref},z} + f^{T_1}_zf^{T_1}_{\text{ref},x}f^{T_1}_{\text{ref},y}) \\
G^{T_2}_1 &= \langle\boldsymbol{f}^{T_2},\boldsymbol{f}^{T_2}\rangle \\
G^{T_2}_2 &= \langle\boldsymbol{f}^{T_2},\boldsymbol{f}^{T_2}_\text{ref}\rangle \\
G^{T_2}_3 &= f^{T_2}_xf^{T_2}_{\text{ref},y}f^{T_2}_{\text{ref},z} + f^{T_2}_yf^{T_2}_{\text{ref},x}f^{T_2}_{\text{ref},z} + f^{T_2}_zf^{T_2}_{\text{ref},x}f^{T_2}_{\text{ref},y} 
\end{align*}

\end{widetext}

\end{document}